{

\documentclass[12pt,a4paper]{article}
\usepackage{graphicx}
\usepackage[T1]{fontenc}
\usepackage[utf8]{inputenc}
\usepackage{textcomp}
\usepackage[sc,osf]{mathpazo}
\usepackage{a4wide}
\usepackage{latexsym,amsthm,amsfonts,amsmath,mathrsfs,amssymb}
\usepackage{dsfont}
\usepackage{accents}
\usepackage[nosort]{cite}
\usepackage{booktabs} 
\usepackage[unicode,implicit]{hyperref}
\usepackage[normalem]{ulem}
\hypersetup{%
  pdftitle    = {Noether-Wald charge in supergravity: the fermionic contribution}
  pdfkeywords = {black holes, thermodynamics, entropy, Wald entropy, gauge
    symmetry, supersymmetry, supergravity, Killing vectors, Killing
    supervectors, Killign spinors, Noether}
   pdfauthor   = {Igor Bandos and Tom\'as Ort\'{\i}n},
  plainpages  = true,
  colorlinks  = true,
  citecolor   = blue,
  urlcolor    = red,
  linkcolor   = black
}
\newcommand{\hepth}[1]{{\tt
\href{http://www.arXiv.org/abs/hep-th/#1}{hep-th/#1}}}
\newcommand{\grqc}[1]{{\tt
\href{http://www.arXiv.org/abs/gr-qc/#1}{gr-qc/#1}}}

\newcommand{\arxiv}[1]{{\tt arXiv:\href{http://www.arXiv.org/abs/#1}{#1}}}
\newcommand{\doi}[1]{{\tt DOI:\href{http://dx.doi.org/#1}{#1}}}

\allowdisplaybreaks

\makeatletter
\@addtoreset{equation}{section}
\makeatother

\pagestyle{empty}

\begin{document}

\begin{flushright}
\small
IFT-UAM/CSIC-23-013\\
May 16\textsuperscript{th}, 2023\\
\normalsize
\end{flushright}

\vspace{1cm}

\begin{center}

  {\Large {\bf Noether-Wald charge in supergravity: the fermionic contribution}}

\vspace{1.2cm}

\renewcommand{\thefootnote}{\alph{footnote}}

{\sl\large Igor Bandos$^{1,2,}$}\footnote{Email: {\tt Igor.Bandos[at]ehu.eus}}
{\sl\large  and Tom\'{a}s Ort\'{\i}n,}$^{3,}$\footnote{Email: {\tt Tomas.Ortin[at]csic.es}}

\setcounter{footnote}{0}
\renewcommand{\thefootnote}{\arabic{footnote}}
\vspace{.6cm}

${}^{1}${\it Department of Physics and EHU Quantum Center,
  University of the Basque Country UPV/EHU, P.O. Box 644, 48080 Bilbao, Spain}

\vspace{0.2cm}

${}^{2}${\it IKERBASQUE, Basque Foundation for Science, 48011, Bilbao, Spain}

\vspace{0.2cm}

${}^{3}${\it Instituto de F\'{\i}sica Te\'orica UAM/CSIC\\
C/ Nicol\'as Cabrera, 13--15,  C.U.~Cantoblanco, E-28049 Madrid, Spain}

\vspace{1cm}


{\bf Abstract}
\end{center}
\begin{quotation}
  {\small We study the invariance of $\mathcal{N}=1,d=4$ supergravity solutions under
    diffeormophisms and show that, in order to obtain consistent conditions
    (``Killing equations'') invariant under local supersymmetry
    transformations, one has to perform supersymmetry transformations
    generated by the superpartner of the vector that generates standard
    diffeomorphisms, just as a superspace analysis indicates.  Using these
    transformations, we construct a Noether-Wald charge of $\mathcal{N}=1,d=4$
    supergravity with fermionic contributions which is diff- Lorentz- and
    supersymmetry-invariant (up to a total derivative).
  }
\end{quotation}

\newpage
\pagestyle{plain}



\section{Introduction}

The determination of the gauge parameters whose associated transformations
leave all the fields of a given configuration invariant (\textit{reducibility}
or \textit{Killing} parameters) is a basic ingredient in the characterization
of the solutions of a given theory and in the definition of their conserved
charges \cite{Barnich:2001jy}. Since Wald's approach to the first law of
black-hole thermodynamics is based on the properties of the conserved charge
associated to the invariance under diffeomorphisms (the so-called
\textit{Noether-Wald charge}) \cite{Lee:1990nz}, the definition and properties
of the Killing parameters is also a fundamental in that context.

In diff-invariant theories with a spacetime metric, the parameters must be
Killing vectors but, in presence of other gauge symmetries, one has to take
into account the gauge transformations induced by the diffeomorphism
\cite{BergmannFlaherty1978}.\footnote{This fact emerged earlier in the
  construction of the so-called \textit{spinorial Lie derivative}
  \cite{kn:Lich} later extended to \textit{Lie-Lorentz derivative}
  \cite{Ortin:2002qb}.}  Crucially, the combined diff-gauge transformations
(expressed through covariant Lie derivatives) leave invariant the fields in a
gauge-covariant fashion. In Wald's approach, the induced gauge transformations
give rise to the work terms in the first law of black-hole mechanics
\cite{Prabhu:2015vua,Elgood:2020svt}.\footnote{The same is true in other
  approaches \cite{Heusler:1993cj}.}

In most theories the spacetime metric is a pure tensor under diffeomorphisms
and its transformation does not induce any gauge transformations: the vector
fields that generate the isometry are standard Killing vectors. In theories
with local supersymmetry, though, the metric is no longer a pure tensor and it
is necessary to take into account the local supersymmetry transformations
induced by the diffeomorphism: the generator will not satisfy the standard
Killing vector equation, but a supercovariant generalization \cite{Vandyck}.
The superpartner of this equation (in pure $\mathcal{N}=1,d=4$ supergravity)
will express the invariance of the gravitino and must take into account the
induced local Lorentz and supersymmetry transformations \cite{Vandyck}. The
induced supersymmetry transformation parameters will give rise to additional,
fermionic, terms in the Noether-Wald charge with potential implications for
the thermodynamics of the black-hole solutions of these
theories.\footnote{The results of \cite{Aneesh:2020fcr} were
  obtained ignoring the local supersymmetry transformations. The resulting
  Noether-Wald charge is that of Einstein's gravity, which is not
  supersymmetry-invariant, as we are going to see.}

The superspace analysis of the problem\footnote{See the complementary
  material. } indicates that these two equations expressing the invariance of
the metric and gravitino must involve a superpartner of the Killing vector
that we can call \textit{Killing spinor}, even though the equation it
satisfies is not the standard one and our spinor is fermionic. Thus, one is
lead to consider superdiffeomorphisms generated by a Killing supervector
superfield whose leading components are the Killing vector and
spinor.\footnote{Earlier work on Killing supervectors was carried out in
 \cite{Buchbinder:1995uq,Kuzenko:2012vd,Kuzenko:2015lca,Howe:2015bdd,Howe:2018lwu,Kuzenko:2019tys,Chandia:2022uyy}.}  Their equations transform
covariantly under all the symmetries of the theory and Killing parameters
remain Killing after any of those transformations.

As a result, the Noether-Wald charge will also be invariant, up to toal
derivatives, under all the local symmetries of the theory. This is an
essential, \textit{sine qua non}, property of the Noether-Wald charge.  All
the modifications from the standard result (the Komar charge) vanish for
vanishing fermionic fields and, therefore, they do not affect the
thermodynamics of purely bosonic black-hole solutions but they may have
interesting consequences for black holes with non-trivial fermionic fields, as
we will discuss in Section~\ref{eq:discussion}. We start by reviewing the
action and equations of motion of $\mathcal{N}=1,d=4$ supergravity in
Section~\ref{sec-N1d4SUGRA}, and stuyding its symmetries and associated
conserved charges (including the Noether-Wald charge) in
Section~\ref{sec-symmetriesandeom}.

\section{$\mathcal{N}=1,d=4$ supergravity}
\label{sec-N1d4SUGRA}

The 1\textsuperscript{st}-order action of $\mathcal{N}=1,d=4$ supergravity
\cite{Freedman:1976xh,Deser:1976eh} in the conventions of \cite{Ortin:2015hya}
is \cite{Deser:1976eh}\footnote{Actually, the action in \cite{Freedman:1976xh} was
  written in the 2\textsuperscript{nd}-order formalism in which it is very
  difficult to check the invariance of the complete action. Notice also that
  the action (\ref{eq:n1d4sugraaction1st}) can be obtained from the earlier
  action of \cite{Volkov:1973jd} by setting to zero the
  Goldstone fermion there; see \cite{Kuzenko:2021sts} for a recent
  discussion. }

\begin{equation}
\label{eq:n1d4sugraaction1st}
  S[e^{a},\omega^{ab},\psi]
  =
  \frac{1}{16\pi G_{N}^{(4)}}
  \int
  \left\{ -\star (e^{a}\wedge e^{b})\wedge R_{ab}
    +2\bar{\psi}\wedge \gamma_{5} \gamma\wedge \mathcal{D}\psi
  \right\}
  \equiv
  \int
  \mathbf{L}\,,
\end{equation}

\noindent
where

\begin{equation}
  e^{a} = e^{a}{}_{\mu}dx^{\mu}\,,
  \hspace{.5cm}
  \psi = \psi_{\mu}dx^{\mu}\,,
  \hspace{.5cm}
  \omega^{ab} = \omega_{\mu}{}^{ab}dx^{\mu} =-\omega^{ba}\,,
  \hspace{.5cm}
  \gamma \equiv \gamma_{a} e^{a}\,,
\end{equation}

\noindent
are the Vierbein, gravitino, spin connection and gamma matrix 1-forms,
respectively, and $\mathcal{D}$ is the (exterior) Lorentz-covariant derivative

\begin{subequations}
  \begin{align}
    \mathcal{D}e^{a}
    & \equiv
      de^{a} -\omega^{a}{}_{b}\wedge e^{b}\,,
    \\
    & \nonumber \\
    \mathcal{D}\psi
    & \equiv
      d\psi -\tfrac{1}{4}\!\!\not\!\omega\wedge \psi
      \equiv
      2\Psi\,,
  \,\,\,\,\,
  \not\!\omega
  \equiv
  \omega^{ab}\gamma_{ab}\,.
  \end{align}
\end{subequations}

$\Psi$ is the gravitino field strength 2-form.
The curvature and torsion 2-forms $R^{ab},T^{a}$ can be defined through the
identities

\begin{equation}
    \mathcal{D}\mathcal{D}e^{a}
     =
     -R^{a}{}_{b}\wedge e^{b}\,,
     \hspace{1cm}
    \mathcal{D}e^{a}
     =
      -T^{a}\,.
\end{equation}

The \textit{total} (Lorentz and general) covariant derivative is $\nabla$
and it satisfies the first Vierbein postulate

\begin{equation}
\nabla e^{a} =  \mathcal{D}e^{a} -\Gamma_{\mu\nu}{}^{a}dx^{\mu}\wedge dx^{\nu}=0\,.
\end{equation}

The solution to the equation of motion of the spin connection is

\begin{subequations}
  \begin{align}
\label{eq:superconnection}
\omega_{abc}(e,\psi)
& =
-\Omega_{abc} +\Omega_{bca} -\Omega_{cab}
=
\omega_{abc}(e)+K_{abc}\,,
    \\
    & \nonumber \\
  \Omega_{\mu\nu}{}^{a}
 & \equiv
 \partial_{[\mu}e^{a}{}_{\nu]} +\tfrac{1}{2} T_{\mu\nu}{}^{a}\,,
    \\
    & \nonumber \\
  \label{eq:torsion}
  T^{a}
  & =
   \tfrac{i}{2}\bar{\psi}\wedge \gamma^{a}\psi\,,
  \end{align}
\end{subequations}

\noindent
where $\omega(e)$ is Levi-Civita spin connection and
$K^{ab}=K_{\mu}{}^{ab}dx^{\mu}$ is the contorsion 1-form.

Using this solution in Eq.~(\ref{eq:n1d4sugraaction1st}) we obtain the
1.5-order action $S[e^{a},\omega^{ab}(e,\psi),\psi]$
\cite{VanNieuwenhuizen:1981ae}. Its variation is given by\footnote{We suppress
  the factor in front of the action in intermediate steps.}

\begin{equation}
  \label{eq:generalvariation}
  \delta S[e^{a},\omega^{ab}(e,\psi),\psi]
  =
  \int
  \left\{\mathbf{E}_{a}\wedge \delta e^{a}
    +\delta\bar{\psi}\wedge \mathbf{E}
    +d\mathbf{\Theta}(\varphi,\delta\varphi)
  \right\}\,,
\end{equation}

\noindent
where

\begin{subequations}
  \begin{align}
    \label{eq:Ea}
    \mathbf{E}_{a}
    & =
      \imath_{a}\star (e^{b}\wedge e^{c})\wedge R_{bc}
      +2\bar{\psi}\wedge \gamma_{5}\gamma_{a}\mathcal{D}\psi\,,
    \\
    & \nonumber \\
    \label{eq:E}
    \mathbf{E}
    & =
      4\gamma_{5}\gamma\wedge \mathcal{D}\psi
      \,,
    \\
    & \nonumber \\
    \label{eq:Theta}
    \mathbf{\Theta}
    & =
      -\star (e^{a}\wedge e^{b})\wedge \delta\omega_{ab}
      +2\bar{\psi}\wedge \gamma_{5}\gamma\wedge \delta\psi\,,
  \end{align}
\end{subequations}

\noindent
are, respectively, the Vierbein and gravitino equations of motion and the
pre-symplectic 3-form. We have simplified $\mathbf{E}$ using the Fierz
identity $T^{a}\wedge \gamma_{5}\gamma_{a}\psi= 0$

\section{Symmetries and conserved charges}
\label{sec-symmetriesandeom}

\subsubsection*{Local Lorentz symmetry}

The 1.5-order action is exactly invariant under local Lorentz transformations

\begin{equation}
  \delta_{\sigma} e^{a}
  =
\sigma^{a}{}_{b}e^{b}\,,
\hspace{1cm}
\delta_{\sigma} \psi
=
\tfrac{1}{4}\!\!\not\!\sigma \psi\,,
\end{equation}

\noindent
where $\sigma^{ab}=-\sigma^{ba}$. Using the associated Noether identity

\begin{equation}
  \label{eq:localLorentzNoether}
    \mathbf{E}_{a}\wedge \delta_{\sigma} e^{a}
    +\delta_{\sigma}\bar{\psi}\wedge \mathbf{E}
    =
    \mathbf{E}_{a}\wedge e^{b}
    \sigma^{a}{}_{b} -\tfrac{1}{4}\sigma^{ab} \bar{\psi} \wedge \gamma_{ab}
    \mathbf{E}
\mathbf{E}^{[a}\wedge e^{b]}
    -\tfrac{1}{4}\bar{\psi} \wedge \gamma^{ab}
     \mathbf{E}
  =
    0\,,
\end{equation}

\noindent
we find the off-shell conserved current

\begin{equation}
  \label{eq:bigidentity}
  \mathbf{J}[\sigma]
   =
-\star (e^{a}\wedge e^{b})\wedge \mathcal{D}\sigma_{ab}
  +\tfrac{1}{2}\sigma^{ab}\bar{\psi}\wedge \gamma_{5}\gamma\wedge
  \gamma_{ab}\psi
 =
d\mathbf{Q}[\sigma]\,,
\end{equation}

\noindent
where Lorentz charge 2-form $\mathbf{Q}[\sigma]$ is given by

\begin{equation}
  \label{eq:Lorentzcharge2form}
  \mathbf{Q}[\sigma]
  =
 -\star (e^{a}\wedge e^{b})\sigma_{ab}\,.
\end{equation}

Following \cite{Barnich:2001jy} we can construct a
conserved charge for each $\kappa^{ab}_{L}$ generating a Lorentz
transformation that leaves invariant all the fields of a given solution of the
equations of motion, $\delta_{\kappa_{L}}e^{a} = \delta_{\kappa_{L}}\psi=0$,
as the integral over a closed 2-surface $\Sigma$:

\begin{equation}
  \mathcal{Q}(\kappa_{L})
  =
  -\frac{1}{16\pi G_{N}{}^{(4)}}
  \int_{\Sigma} \star (e^{a}\wedge e^{b})\kappa_{L\, ab}\,.
\end{equation}

Observe that the conservation of the Lorentz charge

\begin{equation}
    d\mathbf{Q}[\kappa_{L}]
    =
    \mathbf{J}[\kappa_{L}]
    =
    -\star (e^{a}\wedge e^{b})\wedge \delta_{\kappa_{L}}\omega_{ab}
    +2\bar{\psi}\wedge \gamma_{5}\gamma\wedge \delta_{\kappa_{L}}\psi\,,
\end{equation}

\noindent
only needs the invariance of the gravitino and spin connection. Thus, the
above formula can give non-trivial conserved Lorentz charges even if
$\delta_{\kappa_{L}}e^{a}\neq 0$.

\subsubsection*{Local supersymmetry}

Under the local supersymmetry transformations

\begin{equation}
\label{eq:n1d4psusyrules}
  \delta_{\epsilon} e^{a}
   =
    -i\bar{\epsilon}\gamma^{a}\psi\,,
\hspace{1cm}
  \delta_{\epsilon} \psi
  =
\mathcal{D}\epsilon\,,
\end{equation}

\noindent
the 1.5-order action is only invariant up to a total derivative
\cite{Deser:1976eh}

\begin{equation}
  \label{eq:susytotalderivative}
  \delta_{\epsilon}S[e^{a},\psi]
  =
  \int d\left[
    -\star (e^{a}\wedge e^{b})\wedge \delta_{\epsilon}\omega^{ab}
    +2\bar{\epsilon}\gamma_{5}\gamma_{\wedge}\mathcal{D} \psi
\right]\,.
\end{equation}

Using the associated Noether identity \cite{Deser:1976eh,Ortin:2015hya}

\begin{equation}
  \label{eq:susynoetheridentity}
i\gamma^{a}\mathbf{E}_{a}\wedge \psi
+\mathcal{D}\mathbf{E}
=
0\,,
\end{equation}

\noindent
we find the supercharge 2-form

\begin{equation}
  \label{eq:supercharge2form}
  \mathbf{Q}[\epsilon]
  =
  -\frac{1}{8\pi G_{N}^{(4)}}\bar{\epsilon}\gamma_{5}\gamma \wedge\psi\,.
\end{equation}

Again \cite{Barnich:2001jy}, we can construct a conserved
supercharge for each local supersymmetry parameter $\kappa_{S}$ leaving
invariant a solution $\{e^{a},\psi\}$ of the equations of motion,
$\delta_{\kappa_{S}}e^{a}=\delta_{\kappa_{S}}\psi=0$, via the integral

\begin{equation}
  \mathcal{Q}(\kappa_{S})
  \equiv
  -\frac{1}{8\pi G_{N}^{(4)}} \int_{\Sigma}
  \bar{\kappa}_{S}\gamma_{5}\gamma \wedge\psi\,.
\end{equation}

This supercharge is conserved if
$d\mathbf{Q}[\kappa_{S}]\doteq 0$,\footnote{We use $\doteq$ for on-shell
  identities.} which, as in the Lorentz charge case, only demands $\kappa_{S}$
to be a \textit{Killing spinor} satisfying the \textit{Killing spinor
  equation} (KSE) $ \delta_{\kappa_{S}}\psi = \mathcal{D}\kappa_{S} = 0$.

The on-shell local supersymmetry algebra acting on all the fields of the
theory is

\begin{equation}
  \label{eq:localSUSYalgebra}
  [\delta_{\epsilon_{1}},\delta_{\epsilon_{2}}]
  =
  -\pounds_{\xi}+\delta_{\sigma} +\delta_{\epsilon}\,,
\end{equation}

\noindent
where the parameters of the Lie derivative and local Lorentz and supersymmetry
transformations are

\begin{equation}
  \label{eq:susyalbegraparameters}
  \xi^{a}
   \equiv
    -i\bar{\epsilon}_{1} \gamma^{a}\epsilon_{2}\,,
\hspace{1cm}
  \sigma^{a}{}_{b}
   \equiv
    \imath_{\xi}\omega^{a}{}_{b}\,,
\hspace{1cm}
  \epsilon
   \equiv
  \imath_{\xi}\psi\,.
\end{equation}

\subsubsection*{Diffeomorphisms}

When the fields under consideration have some kind of gauge freedom,
infinitesimal diffeomorphisms $\delta_{\xi}x^{\mu} = \xi^{\mu}$ induce
(``compensating'') gauge transformations that need to be taken into account.
In this theory

\begin{equation}
\delta_{\xi} = -\pounds_{\xi} +\delta_{\sigma_{\xi}}+\delta_{\epsilon_{\xi}}\,.
\end{equation}

\noindent
$\sigma_{\xi}{}^{ab}$ and $\epsilon_{\xi}$ are $\xi$-related local Lorentz and
supersymmetry parameters which can be determined when $\xi=k$ such that
$\delta_{k}=0$ on all the fields.

It is natural to start with the metric which, in non-supersymmetric theories,
is just a general tensor so that
$\delta_{k}g_{\mu\nu}=-\pounds_{k}g_{\mu\nu}=0$, the Killing vector equation
(KVE) for $k$. However, in supergravity, the metric transforms under
supersymmetry, $\pounds_{k}g_{\mu\nu}=0$ would not be invariant, and,
according to the general prescription, we must write \cite{Vandyck}

\begin{equation}
  \delta_{\xi}g_{\mu\nu}
  =
  -2\nabla_{(\mu}(e)\xi_{\nu)}
  -2i\bar{\epsilon}_{\xi}\gamma_{(\mu}\psi_{\nu)}
  =
    -2\nabla_{(\mu}\xi_{\nu)}
  -2i\left(\bar{\epsilon}_{\xi}-\imath_{\xi}\bar{\psi}\right)
  \gamma_{(\mu}\psi_{\nu)}\,,
\end{equation}

\noindent
where $\nabla_{\mu}(e)$ is the Levi-Civita covariant derivative. The
superspace analysis suggests that $\epsilon_{\xi}$ is related to the
supersymmetric partner of the vector $\xi$, the spinor $\lambda$, by

\begin{equation}
  \label{eq:epsilonk}
\epsilon_{\xi}=\imath_{\xi}\psi -\lambda \equiv \epsilon_{\xi,\lambda}\,.
\end{equation}

When the spinor $\lambda$ is the superpartner of $k$, it will be denoted by
$\kappa$.  Thus, $(k,\kappa)$ satisfy the following generalization of the KVE
\cite{Vandyck}

\begin{equation}
  \label{eq:superKE}
\nabla_{(\mu}k_{\nu)} -i\bar{\kappa}\gamma_{(\mu}\psi_{\nu)}=0\,,
\end{equation}

\noindent
with integrability condition

\begin{equation}
  \label{eq:integrabilitycondition}
  \imath_{k}R^{ab}+\mathcal{D}P_{k,\kappa}{}^{ab}-\delta_{\epsilon_{k}}\omega^{ab}
  =
  0\,,
\end{equation}

\noindent
where we have defined the supersymmetric generalization of the \textit{Lorentz
  momentum map} or \textit{Killing bivector} \cite{Ortin:2002qb} associated to
$(k,\kappa)$ $P_{k,\kappa\, ab}$ by

\begin{equation}
  \label{eq:Lorentzmm}
  P_{k,\kappa\, ab}
  \equiv
  \nabla_{a}k_{b} -i\bar{\kappa}\gamma_{b}\psi_{a}
  =
  P_{k,\kappa\, [ab]}\,,
\end{equation}

\noindent
or by a 1-form equation

\begin{equation}
  \label{cDka=M}
  \mathcal{D}k^{a}+ i\bar{\psi}\gamma^{a}\kappa
  -e^{b} P_{k,\kappa\, b}{}^{a}
  =
  0\,.
\end{equation}

Observe that this equation assumes Eq.~(\ref{eq:superKE}).

According to the general prescription, for the Vierbein

\begin{equation}
  \delta_{\xi,\lambda} e^{a}
  =
  -\pounds_{\xi}e^{a} + \sigma_{\xi,\lambda}{}^{a}{}_{b}e^{b}
  -i\bar{\epsilon}_{\xi,\lambda} \gamma^{a}\psi\,.
\end{equation}

\noindent
By assumption $\delta_{k,\kappa} e^{a}=0$ and we find

\begin{equation}
  \label{eq:sigmak}
  \sigma_{k,\kappa}{}^{ab}
  =
  \imath_{k}\omega^{ab} -P_{k,\kappa}{}^{ab}\,.
\end{equation}

For the gravitino, \cite{Vandyck}

\begin{equation}
  \delta_{\xi,\lambda} \psi
  =
  -\pounds_{\xi}\psi +\tfrac{1}{4}\!\!\not\!\sigma_{\xi,\lambda}\psi
  +\mathcal{D}\epsilon_{\xi,\lambda}\,.
\end{equation}

\noindent
Using Eqs.~(\ref{eq:sigmak}) and (\ref{eq:epsilonk}),

\begin{equation}
  \label{eq:deltxipsi}
  \delta_{\xi,\lambda} \psi
  =
  -2\imath_{\xi}\Psi-\tfrac{1}{4}\!\!\not\!P_{\xi,\lambda}\psi -\mathcal{D}\lambda\,.
\end{equation}

\noindent
Since, by assumption, $\delta_{k,\kappa} \psi=0$, it follows that $\kappa$
must satisfy the \textit{supersymmetry momentum map} equation that generalizes
the KSE $\delta_{\kappa}\psi=\mathcal{D}\kappa=0$

\begin{equation}
  \label{eq:SUSYmomentummapep}
  2\imath_{k}\Psi+\tfrac{1}{4}\!\!\not\!P_{k,\kappa}\psi +\mathcal{D}\kappa
  =
  0\,.
\end{equation}

\subsubsection*{Covariance of the Killing equations}

The main goal of this paper is to obtain results compatible with all the
symmetries of the theory.  Thus, we must prove that the equations
$\delta_{k,\kappa}g_{\mu\nu}= 0$ (\ref{eq:superKE}) and
$\delta_{k,\kappa}\psi =0$ (\ref{eq:SUSYmomentummapep}) transform (super-)
covariantly under local supersymmetry transformations. The superspace analysis
suggests that $k$ and $\kappa$ transform under supersymmetry as

\begin{equation}
    \delta_{\epsilon}k^{a}
     =
     -i\bar{\epsilon}\gamma^{a}\kappa\,,
     \hspace{1cm}
    \delta_{\epsilon}\kappa
     =
     -\tfrac{1}{4}\!\!\not\!P_{k,\kappa}\epsilon\,.
\end{equation}

It can be shown that the local on-shell supersymmetry algebra
Eqs.~(\ref{eq:localSUSYalgebra}), (\ref{eq:susyalbegraparameters}) also holds
on $k$ and $\kappa$ with these supersymmetry transformations. An important
intermediate result needed to prove this is

\begin{equation}
  \label{eq:deltaepsilonP}
    \delta_{\epsilon}P_{k,\kappa\,ab}
     \doteq
    2i\bar{\epsilon}\gamma_{[a|}\left(\mathcal{D}_{|b]}\kappa
      +\tfrac{1}{4}\!\!\not\!P_{k,\kappa}\psi_{|b]}\right) \,,
\end{equation}

\noindent
or in the equivalent form that follows naturally from the on-shell superfield
formalism

\begin{equation}
  \label{susyPab=}
  \delta_\epsilon P_{k,\kappa\, ab}
  =
  - 4i\bar{\epsilon}\gamma_{[a}(\imath_{k}\Psi)_{b]}\,.
\end{equation}


Using this result one gets a simple and transparent supersymmetry
transformation of the Lorentz momentum map equation in the form
(\ref{cDka=M}):

\begin{equation}
\label{susy=cDka+}
\delta_{\epsilon}
\left(
  \mathcal{D}k^{a}
  +i\bar{\psi}\gamma^{a}\kappa-e^{b} P_{k,\kappa\, b}{}^{a}\right)
=
i  \bar{\epsilon}\gamma^{a}
\left( \mathcal{D}\kappa+2\imath_{k} \bar{\Psi}
  +\tfrac{1}{4}\!\!\not\!P_{k,\kappa}\psi\right) \; .
\end{equation}

For the supersymmetry momentum map equation Eq.~(\ref{eq:SUSYmomentummapep})
one finds

\begin{equation}
  \delta_{\epsilon}  \left(\mathcal{D}\kappa
    +\tfrac{1}{4}\!\!\not\!P_{k,\kappa}\psi
    +2\imath_{k}\Psi\right)
  \doteq
  -\tfrac{1}{4}
  \left(\mathcal{D}P_{k,\kappa\, ab}+\imath_{k}R_{ab}
    -\delta_{\epsilon_{k,\kappa}}\omega_{ab}\right)\gamma^{ab}\epsilon\,.
\end{equation}

\subsubsection*{The Noether-Wald charge}

The Noether-Wald charge is the 2-form associated to the invariance under
diffeomorphisms. In order to get consistent results, we must use the
$\delta_{\xi,\lambda}$ transformations constructed in the previous section
with compensating Lorentz and supersymmetry transformations. The later and the
standard diffeomorphisms only leave the action invariant under diffeomorphisms
up to a total derivative. Thus, we must take into account a total derivative
term of the form

\begin{equation}
  \delta_{\xi,\lambda}S
  =
  -\int \left\{
    \imath_{\xi}\mathbf{L}
    +\star (e^{a}\wedge e^{b})\wedge \delta_{\epsilon_{\xi,\lambda}}\omega^{ab}
    -2\bar{\epsilon}_{\xi,\lambda}\gamma_{5}\gamma\wedge\mathcal{D} \psi
  \right\}\,.
\end{equation}

On the other hand, if we use the general variation of the action
Eq.~(\ref{eq:generalvariation}), we arrive to the identity

\begin{equation}
  \begin{aligned}
    \int \left\{\mathbf{E}_{a}\wedge \delta_{\xi,\lambda} e^{a}
      +\delta_{\xi,\lambda}\bar{\psi}\wedge \mathbf{E}
      +d\left[-\star (e^{a}\wedge e^{b})\wedge \left(\delta_{\xi,\lambda}\omega_{ab}
          -\delta_{\epsilon_{\xi,\lambda}}\omega^{ab}\right)
        \right.\right.
    & \\
    & \\
    \left.    \left.
      +2\bar{\psi}\wedge \gamma_{5}\gamma\wedge \delta_{\xi,\lambda}\psi
      +\imath_{\xi}\mathbf{L}
      -2\bar{\epsilon}_{\xi,\lambda}\gamma_{5}\gamma\wedge\mathcal{D} \psi\right]
  \right\}
  & =
  0\,.
\end{aligned}
\end{equation}
Using the Noether identity associated to local Lorentz invariance
Eq.~(\ref{eq:localLorentzNoether}), the Noether identity associated to diff
invariance:

\begin{equation}
    \mathcal{D}\mathbf{E}_{a}
    +2 \imath_{a}\bar{\Psi}\wedge\mathbf{E}
    =
    0\,,
\end{equation}

\noindent
and the supersymmetry Noether identity Eq.~(\ref{eq:susynoetheridentity}), we
are lead to the off-shell converved current

\begin{equation}
  \label{eq:Noether-Waldcurrent}
  \mathbf{J}[\xi,\lambda]
  =
  \star (e^{a}\wedge e^{b})\wedge \mathcal{D}P_{\xi\, ab}
  -\tfrac{1}{2}\bar{\psi}\wedge \gamma_{5}\gamma\wedge
  \!\!\not\!P_{\xi}\psi
  -2\bar{\lambda} \gamma_{5}\gamma\wedge\mathcal{D} \psi
  +2\mathcal{D}\bar{\lambda}\wedge \gamma_{5}\gamma\wedge\psi\,.
\end{equation}

This current is a total derivative as can be seen by using
Eq.~(\ref{eq:bigidentity}) with $\sigma^{ab}$ replaced by $-P_{\xi}{}^{ab}$
and using the Fierz identities. The result is

\begin{equation}
  \mathbf{J}[\xi,\lambda]
  =
  d\mathbf{Q}[\xi,\lambda]\,,
  \,\,\,\,
  \text{where}
  \,\,\,\,
  \mathbf{Q}[\xi,\lambda]
  =
  \star (e^{a}\wedge e^{b})P_{\xi\, ab}
  +2\bar{\lambda}\gamma_{5}\gamma\wedge \psi
\end{equation}

\noindent
is the Noether-Wald charge 2-form we were after. Indeed, it is manifestly
invariant under diffeomorphisms and local Lorentz
transformations. Furthermore, for Killing parameters $(k,\kappa)$
it can be shown that it is on-shell closed and supersymmetry-invariant up to a
total derivative

\begin{subequations}
  \begin{align}
    d\mathbf{Q}[k,\kappa]
    & \doteq
      0\,,
    \\
    & \nonumber \\
\delta_{\epsilon}\mathbf{Q}[k,\kappa]
& \doteq
d\left(  -2\bar{\kappa}\gamma_{5}\gamma\epsilon  \right)\,.
  \end{align}
\end{subequations}

$\mathbf{Q}[k,\kappa]$ is, thus, the supersymmetric generalization of the
Komar charge. Observe that this charge contains terms corresponding to the
standard Komar charge (a Lorentz charge 2-form
Eq.~(\ref{eq:Lorentzcharge2form}) for the momentum map) plus terms
corresponding to the supercharge Eq.~(\ref{eq:supercharge2form}) neither of
which is separately invariant under supersymmetry.

\section{Discussion}
\label{eq:discussion}

Our definition of invariance under (super-) diffeomorphisms and the
supersymmetric Noether-Wald charge we have derived from it provide a solid
basis to study the supersymmetric thermodynamics of the black holes of
$\mathcal{N}=1,d=4$ supergravity. Unavoidably, and contrary to the results of
\cite{Aneesh:2020fcr}, when they are present, fermions (gravitini) must
play a role in it if local supersymmetry remains unbroken. As a general rule,
gauge fields with gauge freedoms are expected to contribute to the
Noether-Wald charge and, probably, other kinds of fermions will not
contribute, although this needs to be proven.

In order to develop this supersymmetric thermodynamics (themodynamics in
superspace, actually, because, when fermions are present, bosonic fields such
as the metric have body and soul) many notions of Lorentzian geometry need to
be extended to this realm: how are event horizons defined and characterized in
this setting? (How do particles move in this space?) Do they coincide with
(super-) Killing horizons? If so, can one generalize the definition of surface
gravity to them? etc.

Another important ingredient to make this possible generalization relevant is
the existence of black-hole solutions with gravitino hair. It was conjectured
in \cite{Cordero:1978ud} and proven in \cite{Gueven:1980be} that the only ones
in $\mathcal{N}=1,d=4$ (Poincar\'e) supergravity are those whose gravitini can
be removed by a supersymmetry transformation as in \cite{Baaklini:1977bx}. In
 $\mathcal{N}=2,d=4$ (Poincar\'e) supergravity things are different
\cite{Aichelburg:1981rd,Gueven:1982tk} and a solution with non-pure-gauge
gravitini built over the body of the extremal Reissner-Nordstr\"om black hole
exists \cite{Aichelburg:1983ux}. This solution provides a testing ground for
the ideas presented here \cite{kn:BMO}.

The work presented here can obviously be extended to $\mathcal{N}=1,2,d=4$ AdS
supergravity and higher-dimensional generalizations and it can also be applied
to the study of the superalgebras of symmetry of solutions (the prescriptions
used in
\cite{Figueroa-O'Farrill:1999va,Alonso-Alberca:2002wsh,Figueroa-OFarrill:2007omz,Ortin:2015hya}
should follow from our definitions). Work in this direction is currently under
way \cite{kn:BMO}.

\section*{Acknowledgments}

This work has been supported in part by the MCI, AEI, FEDER (UE) grants
PID2021-125700NB-C21 (TO and IB), IFT Centro de Excelencia Severo Ochoa
CEX2020-001007-S (TO) and and by the Basque Government Grant IT1628-22
(IB). TO wishes to thank M.M.~Fern\'andez for her permanent support.

\appendix

In this Appendix  we present some details on Killing supervectors in
  supergravity superspace which inspired the results in the main text,
  obtained in the framework of the spacetime component field formalism.

\section{Killing supervectors in superspace}

\subsection{Notation and notions of the superspace formalism}

To the best of our knowledge, the concept of Killing supervector was
introduced in \cite{Buchbinder:1995uq} and developed in
\cite{Howe:2015bdd}. Recently this problem was addressed in
\cite{Chandia:2022uyy} from the perspective of the pure spinor superstring.

Let us denote the supervielbein 1-form of a supergravity  superspace by

\begin{equation}
  \label{EA=4D1N}
E^{A}=dZ^{M}E_{M}^{A}(Z)=(E^{a},\bar{\mathcal{E}}^{\underline{\alpha}})\; ,
\end{equation}

\noindent
where $a=0,1,\ldots, D-1$ and ${\underline{\alpha}}$ is just the spinor index
when $\mathcal{N}=1$ or an index which combines the spinor and R-symmetry
group indices when $\mathcal{N}>1$.  In the $\mathcal{N}=1$, $D=4$
supergravity case in which we are interested here
${\underline{\alpha}}=1,2,3,4$ is the Majorana spinor index.

In \eqref{EA=4D1N}

\begin{equation}
  Z^{M}= (x^{\mu}, \theta^{\check{\underline{\alpha}}}) \; ,
\end{equation}

\noindent
denotes the local coordinates of the supergravity superspace and the indices
$M=(\mu,\check{\underline{\alpha}})$ are the curved indices of this
superspace.  Notice that the fermionic curved indices
$\check{\underline{\alpha}}$, only can be identified as standard spinor indices
after we use superdiffeomorphisms to fix the Wess-Zumino gauge\footnote{To the
  best of our knowledge the complete set of the Wess-Zumino gauge-fixing
  conditions \eqref{WZgauge} were presented for the first time in
  \cite{Bandos:1985un} following similar considerations for the normal gauge
  in superspace there and in \cite{McArthur:1983fm,Khudaverdian:1982nf}.}

\begin{eqnarray}
  \label{WZgauge}
  \imath_{\theta} E^{A}
  :=
  \theta^{\check{\underline{\alpha}}} E_{\check{\underline{\alpha}}}{}^{A}
  =
  \theta^{\check{\underline{\alpha}}}
  \delta_{\check{\underline{\alpha}}}{}^{A}\; , \hspace{1cm}
  \imath_{\theta}\omega
  :=
  \theta^{\check{\underline{\alpha}}}
  \omega_{\check{\underline{\alpha}}}{}^{ab}=0\; .
\end{eqnarray}

The definitions of differential form and exterior derivatives in superspace
differ form those used in the spacetime component approach of the main
text. In particular, differential forms in superspace are defined as

\begin{equation}
  \label{Omq=}
  \Omega_{q}
  =
  \frac{1}{q!}dZ^{M_{q}}\wedge \cdots\wedge dZ^{M_{1}}\Omega_{M_{1}\ldots M_{q}}(Z)
  =
  \frac{1}{q!}E^{A_{q}}\wedge \cdots\wedge E^{A_{1}}\Omega_{A_{1}\ldots A_{q}}(Z)\,,
\end{equation}

\noindent
and the exterior differential acts from the right, that is

\begin{equation}
  \label{dOmq=}
  \begin{aligned}
    d\Omega_{q}
    & =
    \frac{1}{(q+1)!}dZ^{M_{q}}\wedge \cdots\wedge dZ^{M_{1}}(q+1)
    \partial_{[M_{1}}\Omega_{M_{2}\ldots M_{q+1}\}}(Z)
    \\
    & \\
    & =  \frac{1}{(q+1)!}
    E^{A_{q}}\wedge \cdots\wedge E^{A_{1}}
    (q+1)\nabla_{[A_{1}}\Omega_{A_{1}\ldots A_{q+1}\}}(Z)\,,
  \end{aligned}
\end{equation}

\noindent
(see \cite{Grimm:1978ch} as well as \textit{e.g.}
\cite{Wess:1992cp,Bandos:2002bx} and references therein). The same applies to
the contraction of differential forms

\begin{equation}
  \label{dOmq=}
  \begin{aligned}
    \imath_{\delta} \Omega_{q}
    & =
    \frac{1}{(q-1)!}dZ^{M_{q}}\wedge \cdots\wedge dZ^{M_{2}}\delta Z^{M_{1}}\Omega_{M_{1}M_{2}\ldots M_{q}}(Z)
    \\
    & \\
    & = \frac{1}{(q-1)!}  E^{A_{q}}\wedge \cdots\wedge E^{A_{2}}
    \imath_{\delta} E^{A_{1}} \Omega_{A_{1}A_{2}\ldots A_{q}}(Z)\,.
  \end{aligned}
\end{equation}

These conventions decrease the number of sign factors in supertensor
expressions.

The superspace torsion and curvature 2-forms are defined by \cite{Grimm:1978ch}

\begin{subequations}
  \begin{align}
  T^{A}= DE^{A}
  & =
  dE^{A}-E^{B}\wedge \omega_{B}{}^{A}
  =
  \tfrac{1}{2} E^{C}\wedge E^{B} T_{BC}{}^{A}\; ,
  \\
  & \nonumber \\
  R^{ab}
  & =
    d\omega^{ab}-\omega^{ac}\wedge \omega_{c}{}^{b}
    =
    \tfrac{1}{2}  E^{D}\wedge E^{C} R_{CD}{}^{ab}\; ,
  \end{align}
\end{subequations}

\noindent
where, as indicated above, exterior derivative acts from the right.\footnote{ {Thus, the relation with the notation in the main text and in
    \cite{Ortin:2015hya} is $\omega^{ab}\mapsto -\omega^{ab}$ and
    $R_{cd}{}^{ab}\mapsto -R_{cd}{}^{ab}$, but $T^{a}\mapsto T^{a}$ and
    $R^{ab}\mapsto R^{ab}$}.}

The bosonic and fermionic covariant derivatives
$D_{A}=(D_{a}, \mathfrak{D}_{{\underline{\alpha}}})$ appear in the
decomposition of the covariant differential as

\begin{equation}
  \label{D=ED}
  D
  =
  E^{A}D_{A}
  =
  E^{a}D_{a}
  +\bar{\mathcal{E}}^{\underline{\alpha}}\mathfrak{D}_{\underline{\alpha}}
  =:
  E^{a}D_{a} + \bar{\mathcal{E}}\mathfrak{D}\; .
\end{equation}

\subsection{Killing supervectors in superspace}

Killing supervectors

\begin{equation}
  K^{A}=(K^{a}, \mathcal{K}^{\underline{\alpha}})\; ,
\end{equation}

\noindent
in curved superspace of any bosonic and fermionic dimensions
are defined by the conditions

\begin{eqnarray}
\label{DKA=sKilling}
  -\delta_{K} E^{A}
  & = &
  E^{A}(Z+K)-E^{A}(K)
  =
        DK^{A} + \imath_{K} T^{A} + E^{B} \imath_{K}{\omega}_{B}{}^{A}
        \nonumber \\
  & & \nonumber \\
  & = &
        E^{B} {L}_{B}{}^{A} \; ,
  \\
  & & \nonumber \\
  \label{iKR=sKilling}
  -\delta_{K} \omega^{ab}
  & = &
        \omega^{ab}(Z+K)- \omega^{ab}(K)
        =
        D\imath_{K}\omega^{ab} + \imath_{K} R^{ab}
        \nonumber \\
  & & \nonumber \\
  & = &
    D {L}^{ab} \; ,
\end{eqnarray}

\noindent
where $DK^{A}=dK^{A}-K^{B}\omega_{B}{}^{A}$ and, in the case of
$\mathcal{N}=1$ superspace,

\begin{equation}
  \omega_{B}{}^{C}
  =
  \left(\begin{matrix}\omega_{b}{}^{a} & 0 \\
      & \\
      0 & \tfrac{1}{4}  \omega^{cd} \gamma_{cd}\end{matrix}\right)
  =
  \left(\begin{matrix}\omega_{b}{}^{a} & 0 \\
      & \\
      0 & \tfrac{1}{4}  \omega\!\!\!/{}\end{matrix}\right)\,,
\end{equation}

\noindent
is the spin connection  in the supervector (reducible) representation of Lorentz group
algebra and

\begin{equation}
  L_{B}{}^{A}
  =
  \left(\begin{matrix}L_{b}{}^{a} & 0 \\
      & \\
      0 & \tfrac{1}{4}  L^{cd}
      \gamma_{cd}\end{matrix}\right)
  =
  \left(\begin{matrix}L_{b}{}^{a} & 0 \\
      & \\
      0 & \tfrac{1}{4}
      L\!\!\!/{}\end{matrix}\right)\; .
\end{equation}

\noindent
Of course, in our superspace approach $L^{ab}=L^{ab}(Z)$ is a superfield and
its leading component, is given by $-\sigma^{ab}(x)$ of the main text,

\begin{equation}
L^{ab}\vert_{\theta =0}= -\sigma^{ab}(x)\;.
\end{equation}

In what follows we are going to use the following shorthand notation for the
leading components of superfields:
$(\cdots)\vert_{\theta =0}\equiv (\cdots)\vert$.

We find it convenient to define the following \textit{superspace momentum map}:

\begin{equation}
  P_{(K)B}{}^{A}
  =
  -\imath_{K}{\omega}_{B}{}^{A}+ {L}_{B}{}^{A} (K)
  =
  \left(\begin{matrix}P_{(K)b}{}^{a} & 0 \\
      & \\
      0 & \tfrac{1}{4}  P_{(K)}^{cd} \gamma_{cd}\end{matrix}\right)
  =
  \left(\begin{matrix}P_{(K)b}{}^{a} & 0 \\
      & \\
      0 & \tfrac{1}{4}\!\not\! P_{(K)}\end{matrix}\right)\,,
\end{equation}

\noindent
and write \eqref{DKA=sKilling} and \eqref{iKR=sKilling} as

\begin{subequations}
  \begin{align}
  \label{DKA=sK}
  DK^{A} + \imath_{K} T^{A} -E^{B} P_{(K)B}{}^{A}
  & =
    0\; , \\
  & \nonumber \\
  \label{iKR=sK}
  DP_{(K)}{}^{ab} -\imath_{K} R^{ab}
  & =
    0\; .
  \end{align}
\end{subequations}

\subsubsection{From superfield to component description of Killing supervector I. Generalities on the Wess-Zumino Gauge}

To pass from the superfield to the component formulations of supergravity we
first impose the Wess-Zumino gauge \eqref{WZgauge}

\begin{subequations}
  \begin{align}
  \label{iWZg}
    \imath_{\underline{\theta}} E^{a}
    & =
  0\; , \\
  & \nonumber \\
  \label{iWZgf}
  \imath_{\underline{\theta}} \bar{\mathcal{E}}^{\underline{\alpha}}
  & =
  \bar{\theta}^{\underline{\alpha}} \; ,
  \\
  & \nonumber \\
  \label{iWZw}
  \imath_{\underline{\theta}} w^{ab}
  & =
  0\; .
\end{align}
\end{subequations}

\noindent
These conditions can be written in the equivalent form

\begin{eqnarray}
  \label{WZgD}
  \bar{\theta}^{\underline{\beta}} \; \mathfrak{D}_{\underline{\beta}}
  =:
  \bar{\theta} \mathfrak{D}
  =
  \theta\partial
  :=
  \bar{\theta}^{\check{\underline{\alpha}}} \partial_{\check{\underline{\alpha}}}\; .
\end{eqnarray}

The above equations are valid for any number of spacetime dimensions, hence
the underlined notation ${\underline{\alpha}}$ for spinor indices of fermionic
coordinates. The ``checked'' indices ${\check{\underline{\alpha}}}$ denote the
``world superspace'' indices carried by fermionic coordinates which are
identified with the spinor indices of the Lorentz group after the Wess-Zumino
gauge \eqref{iWZg} is imposed.

It is easy to check that the gauge-fixing conditions \eqref{iWZg} and
\eqref{WZgD} imply

\begin{eqnarray}
  \label{WZ0gg1}
  E_{N}{}^{A}\vert_{\theta =0}
  & = &
  \left(\begin{matrix}e_\nu{}^{a}(x) &
      \bar{\psi}_\nu{}^{\underline{\alpha}}(x)
      \\
      & \\
      0 &
      \delta_{\check{\underline{\beta}}}{}^{\underline{\alpha}} \end{matrix}\right)
          \; ,
  \\
  & & \nonumber \\
  \label{WZ0gg2}
  E_{A}{}^{N}\vert_{\theta =0}
  & = &
\left(\begin{matrix}e_{a}^\nu(x) & -
    \bar{\psi}_{a}{}^{\check{\underline{\beta}}}(x)\\
    & \\
    0 &
\delta_{\underline{\alpha}}{}^{\check{\underline{\beta}}}\end{matrix}\right)
        \; ,
\end{eqnarray}

\noindent
where
$\psi_{a}{}^{\check{\underline{\beta}}}(x)\equiv e_{a}{}^\nu(x)
\psi_\nu{}^{\underline{\alpha}}(x)
\delta_{\underline{\alpha}}{}^{\check{\underline{\beta}}}$.

With \eqref{WZ0gg1} one immediately finds the following relations for the
leading terms of the fermionic and bosonic components of an arbitrary
contravariant supervector
$L_{A}(Z)=(L_{a}, L_{\underline{\alpha}})=E_{A}^{M} L_{M}(Z)$

\begin{eqnarray}
  \label{Ltf0=}
  L_{\check{\underline{\beta}}}\vert
  & = &
  (E_{\check{\underline{\beta}}}^{A}L_{A})\vert
  =
  \delta_{\check{\underline{\beta}}}{}^{\underline{\alpha}}
        L_{\underline{\alpha}}\vert
        \nonumber \\
  & & \nonumber \\
  & = &
        L_{\underline{\beta}}(x) \; ,
  \\
  & & \nonumber \\
  \label{La0=}
  L_{a}\vert
  & = &
  (E_{a}^{N}L_{N})\vert
  =
  e_{a}^{\mu} L_{\mu}(x)
  -\psi_{a}{}^{\check{\underline{\beta}}}(x){}
        L_{\check{\underline{\beta}}}\vert
  \nonumber \\
  & & \nonumber \\
  & = &
  e_{a}^{\mu} L_{\mu}(x)
  -\psi_{a}{}^{\underline{\alpha}} L_{\underline{\alpha}}(x) \; .
\end{eqnarray}

\noindent
In particular, this implies

\begin{equation}
  \label{Df0=}
  (D_{\underline{\alpha}}(\cdots))\vert
  =
  (E_{\underline{\alpha}}^{M} \partial_{M} (\cdots))\vert
  =
  \delta_{\underline{\alpha}}{}^{\check{\underline{\beta}}}
  \partial_{\check{\underline{\beta}}} (\cdots)\vert
  =
  \partial_{\underline{\alpha}} (\cdots)\vert \; ,
\end{equation}

\noindent
so that the leading components of the covariant derivatives of a superfield
will produce the second (next to leading) component in the decomposition on
Grassmann variables of this superfield. The leading component of vector
derivative in superspace is

\begin{equation}
  \label{Da0=}
  (D_{a}(\cdots))\vert
  =
  (E_{a}^{M} \partial_{M} (\cdots))\vert
  =
  e_{a}^{\mu} \partial_{\mu} -\psi_{a}{}^{\check{\underline{\beta}}}
  \partial_{\check{\underline{\beta}}} (\cdots)\vert
  =
  e_{a}^{\mu} \partial_{\mu}  ((\cdots)\vert )
  -\psi_{a}{}^{{\underline{\alpha}}} (D_{\underline{\alpha}}(\cdots))\vert  \; ,
\end{equation}

\noindent
where we have used \eqref{Df0=}.

One can also find that

\begin{eqnarray}
  \label{Tbbf0WZ=Gen}
  T_{ab}{}^{\underline{\alpha}}\vert_{0}
  & = &
        e_{a}^{\mu} e_{b}^{\nu} T_{\mu\nu}{}^{\underline{\alpha}}(x)
        -2\psi_{[a|}{}^{\underline{\beta}}
        T_{\underline{\beta}|b]}{}^{\underline{\alpha}}\vert_{0}
        -\psi_{b}{}^{\underline{\beta}} \psi_{a}{}^{\underline{\gamma}}
        T_{\underline{\gamma}\underline{\beta}}{}^{\underline{\alpha}}\vert_{0}
        \; ,
  \\
  & & \nonumber \\
 \label{Rcdab0WZ=Gen}
  R_{cd}{}^{ab}\vert_{0}
  & = &
        e_{c}^{\mu} e_d^{\nu} R_{\mu\nu}{}^{ab}(x)
        -2\psi_{[c|}{}^{\underline{\alpha}}
        R_{\underline{\alpha}|d]}{}^{ab}\vert_{0}
        -\psi_{d}{}^{\underline{\beta}} \psi_{c}{}^{\underline{\alpha}}
        R_{\underline{\alpha}\underline{\beta}}{}^{ab}\vert_{0}\;,
  \\
  & & \nonumber \\
    \label{Tbbb0WZ=Gen}
    T_{ab}{}^{c}\vert_{0}
    & = &
          e_{a}^{\mu} e_{b}^{\nu} T_{\mu\nu}{}^{c}(x)
          -2\psi_{[a|}{}^{\underline{\beta}} T_{\underline{\beta}|b]}{}^{c}\vert_{0} -  \psi_{b}{}^{\underline{\beta}} \psi_{a}{}^{\underline{\gamma}} T_{\underline{\gamma}\underline{\beta}}{}^{c}\vert_{0}
 \; .
\end{eqnarray}

Similarly to \eqref{La0=} and  \eqref{Ltf0=} we can find

\begin{eqnarray}
  \label{Ka=Ka(x)}
  K^{a}\vert
  & = &
        (K^{N} E_{N}^{a})\vert =K^{\nu} (x)e_\nu^{a}(x) =:k^{a}(x)\; ,
  \\
  & & \nonumber \\
  \label{Kf=KjI+}
  \mathcal{ K}^{\underline{\alpha}}\vert
  & = &
  (K^{N}E_{N}^{\underline{\alpha}})\vert
  =
 \mathcal{ K}^{\check{\underline{\beta}}}(x)
  \delta_{\check{\underline{\beta}}}{}^{\underline{\alpha}}+ K^{\mu}(x)
  \psi_{\mu}^{\underline{\alpha}}(x)
  =:
  \kappa^{\underline{\alpha}}(x)\; .
\end{eqnarray}

When we speak about Killing supervectors in the Wess-Zumino gauge, it is
convenient to introduce a separate notation for the first term in the
r.h.s.~of the above equation

\begin{equation}
  \epsilon_{k,\kappa}{}^{\underline{\alpha}}(x)
  :=
  \mathcal{K}^{\check{\underline{\beta}}}(x)
  \delta_{\check{\underline{\beta}}}{}^{\underline{\alpha}}\,,
\end{equation}

\noindent
so that \eqref{Kf=KjI+} reads

\begin{equation}
  \label{Kf=kf+}
  \kappa^{\underline{\alpha}}(x)
  =
  \epsilon_{k,\kappa}{}^{\underline{\alpha}}(x)
  +k^{\mu}(x) \psi_{\mu}{}^{\underline{\alpha}}(x)
  =
  \epsilon_{k,\kappa}{}^{\underline{\alpha}}(x)
  +\imath_k\psi{}^{\underline{\alpha}}(x)\; .
\end{equation}

Notice that the supersymmetry transformations are well defined for
$k^{a}$ and $\kappa^{\underline{\alpha}}$,

\begin{equation}
  \label{susyK=}
  \delta_{\epsilon} k^{a}(x)
  =
  \epsilon^{\underline{\alpha}} (D_{\underline{\alpha}}K^{a})\vert \; ,
  \qquad
  \delta_{\epsilon} \kappa^{\underline{\alpha}}(x)
  =
  \epsilon^{\underline{\beta}} (D_{\underline{\beta}}\mathcal{K}{}^{\underline{\alpha}})\vert \; .
\end{equation}

\subsection{ Killing supervectors of the on--shell superfield supergravity $\mathcal{N}=1$, $D=4$ }

\subsubsection{On shell superfield supergravity in superspace and spacetime
  supersymmetry on the mass shell }

For simple supergravity on the mass shell, \textit{i.e.}~for the background
solving the ``free'' supergravity equation without source from matter fields
or branes, the superspace torsion 2-forms are

\begin{eqnarray}
  \label{TA=on-shell}
  T^{a}
  & = &
  DE^{a}
  =
   \tfrac{i}{2} \bar{\mathcal{E}} \wedge \gamma^{a}  \mathcal{E}\; ,
   \qquad
  \\
    & & \nonumber \\
  \label{Tal=on-shell}
 \mathcal{T}
   &= &
   D\mathcal{E}
   =
   \tfrac{1}{2} E^{b} \wedge E^{a} \mathcal{T}_{ab}\; ,
   \qquad
\end{eqnarray}

\noindent
and  the curvature of spin connection 1-superform is

\begin{equation}
  \label{4Rab=on-shell}
  R^{ab}
  =
   -i\bar{\mathcal{E}}\wedge \gamma\,{\mathcal{T}}^{ab}
        +\tfrac{1}{2} E^{d}\wedge E^{c} R_{cd}{}^{ab}\; ,
\end{equation}

\noindent
where now $\gamma := E^{a}\gamma_{a}$.

The fermionic torsion superfield which provides the superfield generalization
of the gravitino field strength,

\begin{equation}
  \mathcal{T}_{ab}\vert
  =
  2{\Psi}_{ab}
  =
  2\mathcal{D}_{[a}\psi_{b]}\; ,
\end{equation}

\noindent
is related to the superfield generalization of the Riemann tensor by

\begin{equation}
  \label{DTcd=M}
  \mathfrak{D}\bar{\mathcal{ T}}_{cd}
  =
  -\tfrac{1}{4}\!\!  \not\!\!R_{cd} \; ,
  \qquad
  \not\!\!R_{cd}=  R_{cd}{}^{ab} \gamma_{ab}\; .
\end{equation}

\noindent
The Bianchi identities for the superspace torsion 2-forms

\begin{equation}
  \label{DcT=cER}
  DT^{a}
  =
  -E^{b}\wedge R_{b}{}^{a}\; ,
  \qquad
  D\bar{\mathcal{ T}}
  =
  -\tfrac{1}{4}  \bar{\mathcal{E}}\wedge \not\!\!R \; ,
  \qquad
  \not\!\!R=  R^{ab} \gamma_{ab}\; .
\end{equation}

\noindent
imply, besides  \eqref{DTcd=M}, that the fermionic torsion obeys

\begin{equation}
  \label{selfd}
  \bar{\mathcal{ T}}_{ab}
  =
 \tfrac{i}{2} \epsilon_{abcd} \bar{\mathcal{ T}}{}^{cd}\gamma_5 \; ,
  \qquad
\end{equation}

\noindent
and satisfies the superspace generalization of the Rarita-Schwinger (RS) equation

\begin{equation}
  \label{RS=SSP}
  \epsilon^{abcd}\bar{\mathcal{ T}}_{bc}\gamma_{d}
  =
  0\,,
\end{equation}

\noindent
which can be written in the following equivalent forms:

\begin{equation}
  \label{RS->} \bar{\mathcal{ T}}\gamma^{abc}
  =
  0 \; ,
  \qquad
  \bar{\mathcal{ T}}_{ab}\gamma^{ab}
  =
  0 \; ,
  \qquad
  \bar{\mathcal{ T}}_{ab}\gamma^{b}
  =
  0 \; .
\end{equation}

Supersymmetry is realized in superspace as the fermionic part of
superdiffeomorpisms so that Eqs.~\eqref{TA=on-shell}, \eqref{Tal=on-shell} and
\eqref{4Rab=on-shell} allow to immediately find the expression for the
supersymmetry transformations of the graviton, gravitino and spin connection
in the second order formalism (which in this latter case are valid on the mass
shell).  To this end, we use the formal generalization of the covariantized
Lie derivative formula with contractions defined as

\begin{equation}
  \label{i-eps}
  \imath_{\epsilon} E^{a} =0\; ,
  \qquad
  \qquad
  \imath_{\epsilon} \mathcal{E} =\epsilon\;  ,
  \qquad\qquad
\imath_{\epsilon} \omega^{ab} =0\;  ,
\end{equation}

\noindent
and arrive at

\begin{eqnarray}
  \label{susy=e}
 \delta_{\epsilon} e^{a}
  &=&
 -i \bar{\epsilon}\gamma^{a} \psi\; ,
 \qquad \\
  & & \nonumber \\
  \label{susy=psi}
 \delta_{\epsilon}\psi& = &\mathcal{D}\epsilon \; , \qquad
  \\
  & & \nonumber \\
  \label{susy=om}
\delta_{\epsilon} \omega^{ab} &=& 2i \bar{\epsilon}\,\gamma{\Psi}^{ab}\; ,
\end{eqnarray}

The leading component of the Riemann tensor superfield is related to the
spacetime Riemann tensor by (see \eqref{Rcdab0WZ=Gen})

\begin{eqnarray}
 \label{Rcdab0WZ=onSH}
  R_{cd}{}^{ab}\vert_{0}
& = &
\mathcal{R}_{cd}{}^{ab}-4i\bar{\psi}_{[c}\gamma_{d]}\Psi^{ab}\; ,
  \end{eqnarray}

\noindent
where

\begin{equation}
  \mathcal{R}_{cd}{}^{ab}(x)
  =
  e_{c}^{\mu} e_{d}^{\nu} R_{\mu\nu}{}^{ab}(x)\; .
\end{equation}

\noindent
This implies, in particular, that on the shell of gravitino field equations
\eqref{RS=SSP} are

\begin{equation}
    \label{Rasym=0}
    R_{[abcd]}
    =
    \mathcal{R}_{[abcd]}
    =
    0\,,
\end{equation}

\noindent
where at the last stage we have also used the spacetime Bianchi identities

\begin{eqnarray}
  \mathcal{D}T^{a}\vert
  & = &
  i \bar{\psi}\gamma^{a}\wedge\mathcal{D}{\psi}
  =
  e^{b}\wedge \mathcal{R}_{b}{}^{a}\,,
  \nonumber \\
  & & \nonumber \\
  \Rightarrow
  \qquad
  \mathcal{R}_{[abc]}{}^{d}
  & = &
  2i\bar{\psi}_{[b}\gamma^{a}\Psi_{cd]}\,,
\end{eqnarray}

\noindent
and the spacetime RS equations.

The (on-shell) closure of the supersymmetry algebra is guaranteed by its
superspace origin. However, if one wanted to obtain the explicit form of
the commutators, the explicit form of the supersymmetry transformations of the
gravitino field strength can be useful. Eq.~\eqref{DTcd=M} implies that
$\delta_{\epsilon} \mathcal{T}_{ab}(x) = \tfrac{1}{4}
\not\!\!\mathcal{R}{}_{ab}\epsilon +\tfrac{i}{2}
\bar{\psi}_{[a}\gamma_{b]}\mathcal{T}^{cd}(x) \; \gamma_{cd}\epsilon$, or,
equivalently,

\begin{eqnarray}
  \delta_{\epsilon} {\Psi}_{ab}
  & = &
        \tfrac{1}{8}  \not\!\!\mathcal{R}{}_{ab}\epsilon
        +\tfrac{i}{2} \bar{\psi}_{[a}\gamma_{b]}{\Psi}^{cd}\;
        \gamma_{cd}\epsilon\; .
\end{eqnarray}

\subsubsection{Killing supervector and Lorentz momentum map in
  simple  on-shell Poincar\'e supergravity}

In $\mathcal{N}=1$,$D=4$ on-shell supergravity superspace the super-Killing
equation \eqref{DKA=sK} splits into

\begin{eqnarray}
  \label{DKa=}
  DK^{a}
  & = &
        -i \bar{\mathcal{E}}\gamma^{a} \mathcal{K}+E^{b} P_{(K)b}{}^{a}\; ,
        \qquad \,\,\,\, \text{with} \,\,\,\, \qquad
        P_{(K)b}{}^{a}
        =
        -\imath_{K} \omega_{b}{}^{a}+L_{b}{}^{a}\; ,
         \\
  & & \nonumber \\
  & \text{and}& \nonumber \\
  & & \nonumber \\
\label{DKal=}
  D\bar{\mathcal{K}}
  & = &
        -E^{b}K^{a}\mathcal{T}_{ab}
        +\tfrac{1}{4} \mathcal{E}  \not\!P_{(K)}\; ,
    \quad \,\,\,\, \text{with} \,\,\,\,\quad
        \not\!P_{(K)}
        =
P^{ab}_{(K)}\gamma_{ab}\; .
\end{eqnarray}

Using \eqref{D=ED} one finds that Eq.~\eqref{DKa=} implies the superspace
generalization of the Killing equation

\begin{equation}
  \label{DbDKa==}
  D_{b} K^{a}
  =
  P_{(K) b}{}^{a}
  \qquad \Rightarrow \qquad
  D^{(a}K^{b)}=0\;,
\end{equation}

\noindent
and also

\begin{equation}
  \label{DfKa==}
  \mathfrak{D}K^{a}
  =
  -i\gamma^{a}\mathcal{K} \; .
\end{equation}

Eq.~\eqref{DfKa==} means, in particular, that the spinorial superfields
$\mathcal{K}_{\underline{\alpha}}$ are defined by fermionic derivatives of the
Killing vector superfield,

\begin{equation}
\mathcal{K}=\tfrac{i}{4} \gamma_{a}\mathfrak{D}K^{a}\; .
\end{equation}

\noindent
Furthermore, turning to the lowest dimensional superfield components of
Eq.~\eqref{DKal=}, we find that the derivatives of the composite spinor
superfield $\mathcal{K}$ are defined by

\begin{equation}
  \label{DfcK=M}
  \mathfrak{D} \,  \bar{\mathcal{K}}
  =
  \tfrac{1}{4} \not\!P_{(K)}
  =
  \left( \begin{matrix}
      P_{(K)\beta}{}^{\alpha} & 0 \\ \\  0  &
      P_{(K)}{}^{\dot{\beta}}{}_{\dot{\alpha}}
    \end{matrix}\right)\; .
\end{equation}

The second equality uses the Weyl spinor representation and is presented to
show that the restrictions imposed by Eq.~\eqref{DfcK=M} are quite strong.

Now, let us turn to Eq.~\eqref{iKR=sK}, which in our case reads

\begin{equation}
 DP_{(K)}^{ab}
  =
K^{c}\left(E^{d}R_{cd}{}^{ab}-i\bar{\mathcal{E}}\gamma_{c}\mathcal{T}^{ab}\right)
  +i\bar{\mathcal{K}}{\gamma}\mathcal{T}^{ab}\; .
\end{equation}

\noindent
This equation splits into

\begin{subequations}
  \begin{align}
  \label{DdPK=}
  D_{d}P_{(K)}^{ab}
   & =
   K^{c} R_{cd}{}^{ab} +i\bar{\mathcal{K}}{\gamma}_{d}\mathcal{T}^{ab}\;,
    \\
    & \nonumber \\
  \label{DfPK=}
  \mathfrak{D}P_{(K)}^{ab}
  & =
        -iK^{c} {\gamma}_{c}\mathcal{T}^{ab}
        \equiv
        -i\not\!K\, \mathcal{T}^{ab} \; .
  \end{align}
\end{subequations}

\subsubsection{Components of Killing supervector superfields: Killing vector,
  its fermionic superpartner, the momentum map, and the equations for all of these}

The leading component of \eqref{DbDKa==} in the Wess-Zumino gauge gives

\begin{equation}
  \label{Lab=}
  \begin{aligned}
    P_{(k,\kappa )b}{}^{a} (x)
    & =
    K_{b}{}^{a} (x)
    \\
    & \\
    & =
    (D_{b}K^{a})\vert
    \\
    & \\
    & =
    e_{b}^{\mu} D_{\mu} k^{a}(x)+i\bar{\psi}_{b}\gamma^{a}{\kappa}
    \\
& \\
& =
\mathcal{D}_{b} k^{a}(x)+i\bar{\psi}_{b}\gamma^{a}{\kappa}\;,
  \end{aligned}
\end{equation}

\noindent
where

\begin{equation}
  k^{a}(x)
  =
  K^{a}\vert
  :=
  K^{a}(x,0,0)\; ,
  \qquad
  \kappa (x)
  =
  \mathcal{K} \vert  \;,
\end{equation}

\noindent
and

\begin{equation}
  \mathcal{D}_{a}
  :=
  e_{b}^{\mu} D_{\mu}\vert  \; ,
 \qquad
 \mathcal{D}
 =
 e^{a} \mathcal{D}_{a}
 =
 dx^{\mu} D_{\mu}\vert\,,
\end{equation}

\noindent
is the covariant derivative of spacetime fields.

Eq.~\eqref{Lab=} implies the supersymmetric generalization of the Killing
equation\footnote{ {{\it Cf.} (3.17) in the main text.}}

\begin{eqnarray}
  \label{DaKb+=}
  \mathcal{D}^{(b} k^{a)}
  & = &
        -i\bar{\psi}^{(b}\gamma^{a)}{\kappa}
        =
        i\bar{{\kappa}}\gamma^{(a}\psi^{b)} \;.
\end{eqnarray}

However, technically, it is convenient to work with the complete equation
\eqref{Lab=}, containing besides \eqref{DaKb+=} also the definition of the
spacetime momentum map, and to present it in terms of differential forms on
spacetime as

\begin{equation}
  \label{cDka=}
  \mathcal{D}k^{a}
  +i\bar{\psi}\gamma^{a}\kappa
  =
  e^{b} P_{(k,\kappa)b}{}^{a}\; .
\end{equation}

The superpartner of this equation, the leading component of the
$\propto E^{a}$ part of the superform equation \eqref{DKal=}, can also be
presented as a spacetime 1-form equation\footnote{ {{\it Cf.} (3.25) in the main text. }}

\begin{equation}
  \label{cDkap=M}
  \mathcal{D}{\kappa}
  =
        - 2\imath_{k} {\Psi}-\tfrac{1}{4} \,\not\!P_{(k,\kappa)}\,{\psi}\,,
        \,\,\,\qquad
        \imath_{k}\Psi
        =
        e^{b}k^{a} \Psi_{ab}\;.
\end{equation}

Finally, the leading component of the superfield equation
\eqref{DdPK=} reads\footnote{ {{\it Cf.} (3.18) in the main text.}}

\begin{equation}
  \label{cDPab=}
  \mathcal{D}P^{ab}_{(k,\kappa)}
   =
        \imath_{k}\mathcal{R}^{ab}
        +2i (\bar{\kappa} -\imath_{k}\bar{\psi}))\gamma\Psi^{ab}\,,
\end{equation}

\noindent
and implies

\begin{equation}
  \mathcal{D}\!\!\not\!P_{(k,\kappa)}
  =
  \imath_{k}\mathcal{R}_{\beta}{}^{\alpha}
  +2i \gamma_{ab}\gamma ({\kappa} -\imath_{k}{\psi})\; \bar{\Psi}^{ab}\; .
\end{equation}

In order to find \eqref{cDPab=} we have used \eqref{Da0=} and
\eqref{Rcdab0WZ=Gen} which, together with \eqref{4Rab=on-shell}, gives
\eqref{Rcdab0WZ=onSH}.

Notice that Eq. \eqref{cDPab=} can be written in the
form\footnote{{ { {\it Cf.}  (3.18) in the main text.}}}

\begin{equation}
  \label{cDPab==}
  \mathcal{D}P^{ab}_{k,\kappa}
  =
  \imath_{k}\mathcal{R}{}^{ab} + \delta_{\epsilon_{k,\kappa}}\omega^{ab}\,,
\end{equation}

\noindent
where $\delta_{\epsilon}\omega^{ab}$ is defined in \eqref{susy=om}
and\footnote{{ {See \eqref{Kf=kf+}, {\it cf.}  (3.16) in the
      main text.}}}

\begin{equation}
  \label{cDPab==}
  \epsilon_{k,\kappa}
  =
  {\kappa}-\imath_{k}{\psi}\,.
\end{equation}

\subsubsection{Supersymmetry transformations of the Killing vector, its
  fermionic superpartner and of the momentum map}

Thus far we have shown how the Killing supervector equation in the curved
superspace of the on-shell simple supergravity reproduce the Killing equation,
its superpartner and the equation for the Lorentz
momentum.\footnote{ {{\it Cf.}~Eqs. (3.17) with (3.19), (3.25)
    and (3.18) of the main text.}}

Now we are going to show how the supersymmetry transformations of the Killing
vector, its spinorial superpartner (generalized fermionic Killing spinor), and
the momenum map follow from this on-shell superfield formalism.

For the first two,  these are defined by Eqs.~\eqref{susyK=} which in the case
of the superspace of the on-shell $D=4$ supergravity have to be specified with the
use of the contraction $\imath_{{\epsilon}}$ \eqref{i-eps} of Eqs.~ \eqref{DKa=}
and \eqref{DKal=}. Furthermore, in the second case we can use Eq.~\eqref{Lab=}
to specify the contraction of \eqref{DKal=}:

\begin{equation}
  \label{DfcK=M}
  \mathfrak{D} \,  \bar{\mathcal{K}} \vert
  =
  \tfrac{1}{4} \not\!P_{k,\kappa} \vert
  =
  \tfrac{1}{4} \gamma_{ab}\, \left(\mathcal{D}^{[a} k^{b]}
- i \bar{\psi}^{[a}\gamma^{b]}{\kappa}\right)\,.
\end{equation}

\noindent
Thus in the case of $D=4$, $\mathcal{N}=1$ on-shell supergravity, where
\eqref{DfKa==} and \eqref{DfcK=M}  hold, the supersymmetry transformations
\eqref{susyK=} are\footnote{ {{\it Cf.} (3.26) in the main text.}}

\begin{eqnarray}
\label{susyKa=4DonM}
  \delta_{\epsilon} k^{a}(x)
  & = &
        -i\bar{\epsilon}\gamma^{a}{\kappa}\; ,
  \\
  & & \nonumber \\
\label{susybKal=4DonM}
  \delta_{\epsilon} \bar{\kappa}
  & = &
        \tfrac{1}{4} \bar{\epsilon } \not\!P_{k,\kappa}
        \nonumber \\
  & & \nonumber \\
      & = &
 \tfrac{1}{4} \bar{\epsilon } \gamma_{ab}\, \left(\mathcal{D}^{[a} k^{b]}
      - i \bar{\psi}^{[a}\gamma^{b]}{\kappa}\right)\; .
  \\
  & & \nonumber \\
\label{susyKal=4DonM}
      \Leftrightarrow
      \qquad
  \delta_{\epsilon} {\kappa}
  & = &
        - \tfrac{1}{4}  \not\!P_{k,\kappa}  \epsilon
        \nonumber \\
  & & \nonumber \\
  & = &
 -\tfrac{1}{4}  \gamma_{ab}\epsilon\, \left(\mathcal{D}^{[a} k^{b]}
      - i \bar{\psi}^{[a}\gamma^{b]}{\kappa}\right)\; .
\end{eqnarray}

Similarly, Eqs.~\eqref{DfPK=} imply that the supersymmetry variations of
$ P_{k,\kappa}^{ab}$ from \eqref{Lab=} reads (up to the terms proportional to
the equations of motion of gravitino)\footnote{ {{\it Cf.}
    (3.28) of the main text.}}

\begin{equation}
  \label{susyPK=}
  \delta_{\epsilon} P_{k,\kappa}^{ab}
   =
     -2i \bar{\epsilon} k\!\!\!/{}\, {\Psi}^{ab}\,
\end{equation}

Notice also that, using \eqref{susy=om}, this equation can be written in the form

\begin{eqnarray}
  \label{susyPK=-iksusyom}
  \delta_{\epsilon} P_{k,\kappa}^{ab}
  =
  -\imath_{k} \delta_{\epsilon} \omega^{ab}\,.
\end{eqnarray}

To check the consistency of the transformation property \eqref{susyPK=}, we
can substitute it into the supersymmetry variation of \eqref{cDka=} and find, after some algebra and using the gravitino field equations,

\begin{equation}
  \label{susy=cDka+}
  \delta_{\epsilon} \left(\mathcal{D}k^{a}+ i\bar{\psi}\gamma^{a}\kappa
    -e^{b} P_{k,\kappa\, b}{}^{a}\right)
=
+i  \bar{\epsilon}\gamma^{a}
\left( \mathcal{D}\bar{\kappa}+2\imath_{k} \bar{\Psi}+\tfrac{1}{4}
\not\!P_{k,\kappa}{\psi}\right) \; .
\end{equation}

Now, calculating the supersymmetry transformations of the fermionic superpartner \eqref{cDkap=M} of the Killing vector plus momentum map equation \eqref{cDka=} we find (again, after some algebra and with the use of gravitino field equations)

\begin{equation}
  \label{susy=cDkap+}
\delta_{\epsilon}  \left( \mathcal{D}\bar{\kappa}+2\imath_{k} \bar{\Psi}+\tfrac{1}{4}
\not\!P_{k,\kappa}{\psi}\right)
=
\tfrac{1}{4}  \bar{\epsilon}
\left(\mathcal{D}\!\!\not\!P_{k,\kappa} -\imath_{k}\! \not\!\mathcal{R}
  -\delta_{\epsilon_{k,\kappa}}\!\not\!\omega
\right) \; .
\end{equation}

Although it is tempting to state now that the supermultiplet of the Killing spinor
equation includes also the leading component of \eqref{DdPK=},
Eq. \eqref{cDPab=},

\begin{equation}
  \label{cDPkab=}
  \mathcal{D}P_{k,\kappa}^{ab}
   =
 \imath_{k}\mathcal{R}^{ab}+2i(\bar{{\kappa}}-\imath_{k}\bar{\psi}){\gamma}\,{\Psi}^{ab}\;,
\end{equation}

\noindent
this actually can be obtained as selfconsistency conditions of Killing
equation in its form of \eqref{cDka=}, so that we have  the standard
situation of supersymmetry transforming bosonic object (equation) in terms of
fermionic one and fermionic object (equation) in terms of the derivative of
the bosonic one.



\end{document}